\documentclass[twocolumn,3p,times]{elsarticle}

\usepackage{amssymb}
\usepackage{graphics}
\usepackage{graphicx}
\usepackage{txfonts}
\usepackage{epsfig}
\usepackage{bm}   
\usepackage{dcolumn}
\usepackage{rotate}
\usepackage{amsfonts}
\usepackage{longtable}
\usepackage{latexsym,rotating,wrapfig}
\usepackage{footmisc,booktabs,amssymb,longtable}

\usepackage{xcolor,colortbl}

\usepackage{multirow}
\definecolor{Gray}{gray}{0.85}
\definecolor{LightCyan}{rgb}{0.88,1,1}

\usepackage[scaled=0.92]{helvet}
\usepackage[T1]{fontenc}
\usepackage{textcomp}
\usepackage{braket}
\usepackage{textgreek}
\usepackage{rotate}
\usepackage{tabularx}
\usepackage[referable]{threeparttablex}
\usepackage{epigraph}
\begin{document}

\begin{frontmatter}

\title{In Search of Truth: In memory of Balraj Singh}

\author{Jos\'e Nicol\'as Orce}
\address{Department of Physics \& Astronomy, University of the Western Cape, P/B X17, Bellville 7535, South Africa}

\author{Boris Pritychenko}
\address{National Nuclear Data Center, Brookhaven National Laboratory, Upton, NY 11973-5000, USA }

\author{Tibor Kib\'edi}
\address{Department of Nuclear Physics and Accelerator Applications, Research School of Physics, The Australian National University, Canberra, Australian Capital Territory 2601, Australia}

\author{Jun Chen}
\address{Facility for Rare Isotope Beams, Michigan State University, East Lansing, MI 48824, USA}
\end{frontmatter}

\epigraph{“It took less than an hour to make the atoms, a few hundred million years to make the stars and planets, but five billion years to make man!” \dots}{George Gamow in The Creation of the Universe~\cite{gamow1952creation}}
\vspace{-0.5cm}
\begin{figure}[!ht]
\begin{center}
  \includegraphics[width=3.3cm,height=3.8cm,angle=0]{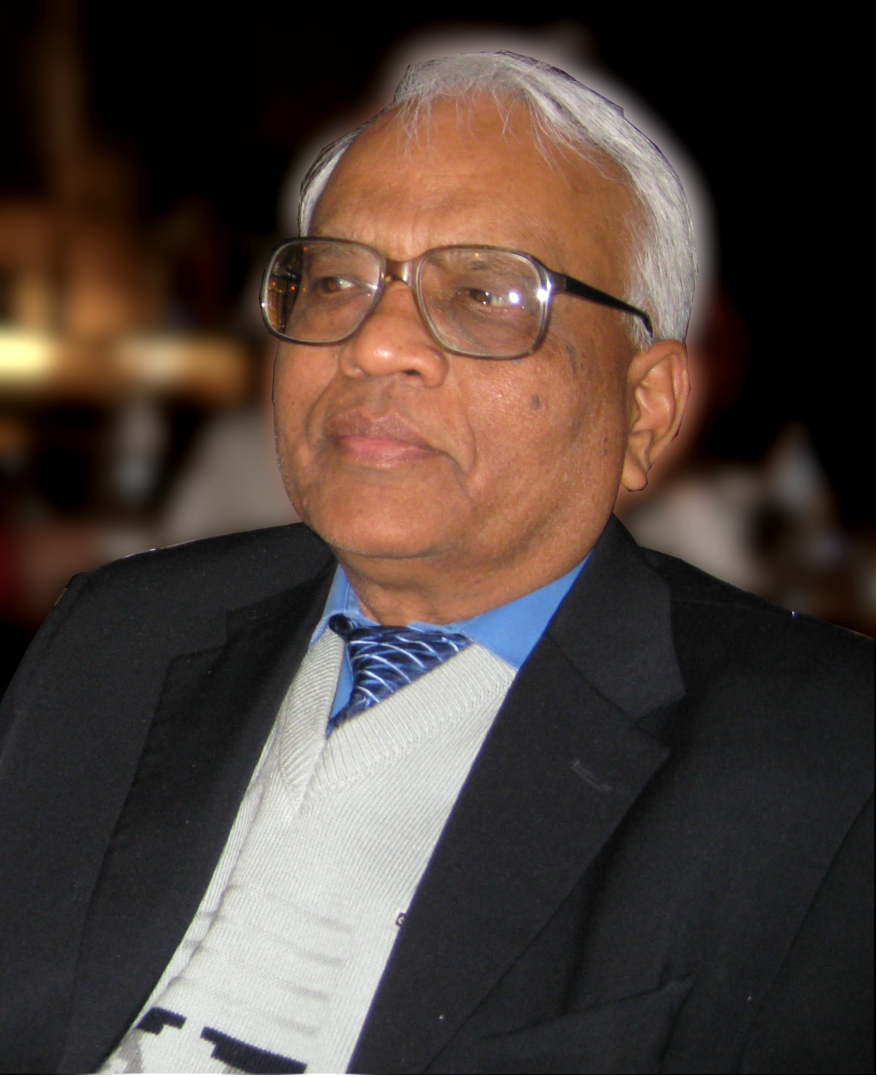}
\label{fig:balraj}
\end{center}
\end{figure}
Born in Punjab (India) in December 1941, Balraj Singh is not only the single most prolific nuclear data evaluator and disseminator of nuclear structure and decay data with 148 evaluations in Nuclear Data Sheets --- 85 as the first and often only author --- plus other journals,
but his upmost curiosity and dedication brought him to be one of the finest nuclear physicists, with an everlasting influence on many of us.
Balraj passed away about a year ago on 9 October 2023 in Ottawa, Ontario (Canada) at the age of 81,
and at Atomic Data and Nuclear Data Tables we would like to commemorate some of his scientific achievements.

We cannot bring up the full dimension of Balraj Singh's persona in one short article. Some of his achievements and curriculum vitae have already been published in Nuclear Data Sheets~\cite{dimitriou2024memory}.
One addition is that he spent one year at the University of Colorado, where he met George Gamow,
who made a strong impression on him.
Behind his immense kindness and modesty, Balraj was an outstanding and meticulous scientist with an inquisitive mind and a deep understanding of
both experimental and theoretical nuclear physics, acquired not only by evaluating data and looking for inconsistencies very critically
---  including changing misprints or incorrect quotations --- but also through understanding the subtle weaknesses of various experimental methods and calculations. For this, he was part of experiments using different spectroscopic probes at different research facilities such as the University of Kentucky in the USA or Chalk River and TRIUMF in Canada.
He was knowledgeable on theoretical models and computer codes,
which altogether made him the ultimate evaluator. Once a new critical result arose, his gentle soul became that of the sharpest detective, a world-trained investigator that would often lead projects from behind until a solution was found.
Balraj was not only a giant in mass-chain evaluation as an evaluator, mentor and also a scientist, but also cared about the {\sc ENSDF}
codes that are also very important in data compilation and evaluation for improving productivity, efficiency and quality,
and made significant contributions by providing ideas, guidance, suggestions/feedbacks.
He never asked for any credit. He never retired.

His hunger for knowledge was additionally fed by his constant interaction with researchers worldwide, whenever true data were concerned. The final result had to make sense. This is how many of us met him the first time. His humble personality made him accessible to anyone and was often approached by colleagues when conflicting data appeared,
supporting researchers from all over the world whenever possible and helping the development of nuclear physics research enormously.
Among many nuclear ``mysteries'' that Balraj undertook, we could mention the discrepancy between Coulomb-excitation and lifetime measurements in the relevant nickel and tin isotopes. The former concerns the lower collectivity or lower
{\sc E2} transition probabilities for the first 2$^+$ states of the neutron-deficient nickel isotopes, which led the development of new JUN45 
shell-model interactions~\cite{honma2009new}  as well as new Coulomb-excitation~\cite{allmond2014high} and lifetime~\cite{orce2008determination,chakraborty2011status} measurements, which are currently tabulated by Pritychenko and collaborators~\cite{pritychenko2016tables}; one of them being, of course, Balraj.
The loss of collectivity in the tin isotopes led to an even greater tumult with long controversy among theoretical approaches trying to explain the
asymmetric behaviour of the {\sc E2} transition probabilities from the first 2$^+$ states~\cite{pritychenko2016tables}, for which he also provided an explanation~\cite{maheshwari2016asymmetric}.
Another interesting case that Balraj helped solving was the anomalous behaviour of the first 2$^+$ excited states in $^{94}$Zr~\cite{elhami2007anomalous},
where the second 2$^+$ state was found to be more collective, which resulted in a better understanding of lifetime measurements~\cite{peters2013level} and the rise of
subshell effects in shape coexistence through $\beta$-decay studies at TRIUMF RIB facility; measurements that he carried out
with colleagues and co-authored in Physical Review Letters~\cite{chakraborty2013collective}.
Additional recent work accounts for the $^{137}$Cs $\beta$-decay in {\sc ENSDF} (with Caroline Nesareja), DDEP (with Sylvain Leblond), the IAEA project to assemble Decay Data Library for Monitoring Applications (with Tibor Kib\'edi) as well as  developing new Java codes ---  e.g., AlphaHF for calculating $\alpha$-decay hindrance factors or KeynunberCheck for checking format and relevance of {\sc NSR} keynumbers cited in {\sc ENSDF}.

Another interesting fact is that his father was a high school principal, so Balraj highly valued education. He had remarkable results with students at McMaster University in Canada, whom he trained in nuclear physics and data evaluation with new techniques that he developed  as well as with students worldwide that he lectured in dedicated workshops ---  e.g., the {\sc ICTP-IAEA} Workshop on Nuclear Structure and Decay Data: Experiment, Theory and Evaluation, held in Trieste (Italy) during 3 -- 14 October 2022 --- and  even became co-authors in some of his leading publications~\cite{balraj2023nuclear}.

Thanks Balraj for all your hard work, trust and encouragement. It has been a great honour to have met you and collaborate with a man of your caliber, from whom we have hopefully
learned to be better researchers and people, always searching for the truth and broader understanding of nature and humanity. Following your lead,
we shall keep your legacy and love for nuclear data to motivate future generations. It is the least we can do after 5 billion years in the making.


\bibliographystyle{elsarticle-num}

\bibliography{balraj}

\end{document}